\documentclass[aps,prd,12pt,superscriptaddress,nofootinbib,floats]{revtex4-2}
\usepackage{bm}
\usepackage{soul}
\usepackage{indentfirst}
\usepackage{amsmath}
\usepackage{graphicx}
\usepackage{float}
\usepackage{amssymb}
\usepackage{subfigure}
\usepackage{amssymb}
\usepackage{psfrag}
\usepackage{hyperref}
\usepackage{epstopdf}
\usepackage{tensor}
\usepackage{multirow}
\usepackage{comment}
\hypersetup{
  colorlinks=true,
  linkcolor=red,
  citecolor=blue,
}

\allowdisplaybreaks[1]

\begin{document}

\title{Scalar induced gravitational waves in light of Pulsar Timing Array data}
\author{Zhu Yi}
\email{yz@bnu.edu.cn}
\affiliation{Advanced Institute of Natural Sciences, Beijing Normal University, Zhuhai 519087, China}
\author{Qing Gao}
\email{gaoqing1024@swu.edu.cn}
\affiliation{School of Physical Science and Technology, Southwest University, Chongqing 400715, China}
\author{Yungui Gong}
\email{yggong@hust.edu.cn}
\affiliation{School of Physics, Huazhong University of Science and Technology, Wuhan, Hubei
430074, China}
\affiliation{Department of Physics, School of Physical Science and Technology, Ningbo University, Ningbo, Zhejiang 315211, China}
\author{Yue Wang}
\email{m202170265@hust.edu.cn}
\affiliation{School of Physics, Huazhong University of Science and Technology, Wuhan, Hubei
430074, China}
\author{Fengge Zhang}
\email{Corresponding author. zhangfg5@mail.sysu.edu.cn}
\affiliation{School of Physics and Astronomy, Sun Yat-sen University, Zhuhai 519088, China}

\begin{abstract}
The power-law parametrization 
for the energy density spectrum of gravitational wave (GW) background is 
a useful tool to study its physics and origin. 
While scalar induced secondary gravitational waves (SIGWs) from some particular models fit the signal detected by NANOGrav, Parkers Pulsar Timing Array, European Pulsar Timing Array, and Chinese Pulsar Timing Array collaborations better than GWs from supermassive black hole binaries (SMBHBs), 
we test the consistency of the data with the infrared part of SIGWs which is somewhat independent of models.
Through Bayesian analysis, 
we show that the infrared parts of SIGWs fit the data better than GW background from SMBHBs.
The results give tentative evidence for SIGWs.
\end{abstract}

\maketitle

\section{Introduction}
The detection of a common-spectrum process, exhibiting the Hellings-Downs angular correlation characteristic of gravitational waves (GWs), 
has been reported by the North American Nanohertz Observatory for Gravitational Waves (NANOGrav) \cite{NANOGrav:2023gor, NANOGrav:2023hde}, 
Parkers Pulsar Timing Array (PPTA) \cite{Zic:2023gta, Reardon:2023gzh},  
European Pulsar Timing Array (EPTA) along with Indian Pulsar Timing Array  (InPTA) \cite{Antoniadis:2023lym, Antoniadis:2023ott}, 
and Chinese Pulsar Timing Array (CPTA) \cite{Xu:2023wog}.  
Estimations based on signals originating from an ensemble of binary supermassive black hole inspirals and a fiducial characteristic-strain spectrum $f^{-2/3}$ suggest a strain amplitude of $2.4^{+0.7}_{-0.6} \times 10^{-15}$ at a reference frequency of $1 ~ \rm{yr}^{-1}$ \cite{NANOGrav:2023gor}. 

While supermassive black hole binaries (SMBHBs) present a plausible explanation for the observed signal \cite{NANOGrav:2023hfp, NANOGrav:2023hvm, Antoniadis:2023zhi,  Ellis:2023dgf, Shen:2023pan, Bi:2023tib, Barausse:2023yrx},  
it is also conceivable that the signal arises from cosmological processes such as cosmic inflation  \cite{NANOGrav:2023hvm, Antoniadis:2023zhi}, first-order phase transitions~\cite{Addazi:2023jvg, Athron:2023mer, Zu:2023olm,Jiang:2023qbm, Xiao:2023dbb,Abe:2023yrw,Gouttenoire:2023bqy,An:2023jxf}, cosmic strings~\cite{Kitajima:2023vre,Ellis:2023tsl,Wang:2023len,Antusch:2023zjk,Ahmed:2023pjl,Ahmed:2023rky,Basilakos:2023xof,Chen:2023zkb}, domain walls~\cite{Kitajima:2023cek,Blasi:2023sej,Babichev:2023pbf}, or scalar-induced gravitational waves (SIGWs) \cite{Franciolini:2023pbf,Liu:2023ymk, Vagnozzi:2023lwo, Cai:2023dls, Wang:2023ost, Jin:2023wri, Liu:2023pau,You:2023rmn,Yi:2023npi,Yi:2023tdk, Bringmann:2023opz,Zhang:2023nrs}. 
In particular, it was shown that SIGWs from three particular models give a better fit than the SMBHB model with Bayes analysis \cite{NANOGrav:2023hvm}.
Since in the infrared region, the  energy density spectrum of SIGWs has a universal log-dependent index for a broad range of models \cite{Yuan:2019wwo,Pi:2020otn},
we would like to ask whether PTAs detected the log-dependent index, or even whether the signal is coming from SIGWs.
Therefore, this paper specifically focuses on investigating the scenario where the signal is attributed to SIGWs.  For other works about this and previous PTA signals, please see \cite{Lu:2023mcz,Guo:2023hyp,Murai:2023gkv,Ghosh:2023aum,Oikonomou:2023qfz,Li:2023yaj,Franciolini:2023wjm,Konoplya:2023fmh, Wu:2023hsa,Wu:2023dnp,Wu:2023pbt,PPTA:2022eul,Chen:2022azo,Chen:2021ncc,Wu:2021kmd,Chen:2021wdo,Chen:2019xse,Ashoorioon:2022raz,Zhang:2023lzt,Zhao:2023joc, InternationalPulsarTimingArray:2023mzf,IPTA:2023ero,Wu:2023rib,Bi:2023ewq,Chen:2023uiz}.

SIGWs, accompanied by the formation of primordial black holes (PBHs),  are sourced by primordial curvature perturbations generated during the inflationary epoch \cite{Hawking:1971ei, Carr:1974nx, Ananda:2006af, Baumann:2007zm, Saito:2008jc,Alabidi:2012ex, Sasaki:2018dmp, Nakama:2016gzw, Kohri:2018awv, Di:2017ndc, Cheng:2018yyr,Cai:2019amo, Cai:2018dig, Cai:2019elf, Cai:2019bmk,Cai:2020fnq, Pi:2020otn, Domenech:2020kqm, Liu:2018ess, Liu:2019rnx,Liu:2020cds,Liu:2021jnw,Liu:2022iuf, Yuan:2019fwv,   Yuan:2019wwo, Yuan:2019udt, Carr:2020gox, Papanikolaou:2021uhe,Papanikolaou:2022hkg,Chakraborty:2022mwu,Chen:2018czv,Chen:2018rzo,Chen:2019irf,Chen:2021nxo,Chen:2022fda, Zheng:2022wqo,Chen:2022qvg, Garcia-Saenz:2023zue, Wu:2020drm, Meng:2022low}. 
These GWs possess a broad frequency distribution and can be detected not only by  PTAs but also by space-based GW detectors, 
such as the Laser Interferometer Space Antenna (LISA) \cite{Danzmann:1997hm,Audley:2017drz}, Taiji \cite{Hu:2017mde}, TianQin \cite{Luo:2015ght,Gong:2021gvw}, 
and the Deci-hertz Interferometer Gravitational-Wave Observatory (DECIGO) \cite{Kawamura:2011zz}. 
The study of SIGWs provides valuable insights into the inflationary period and its dynamics. 
In order to generate significant SIGWs, 
it is expected that the amplitude of the power spectrum of primordial curvature perturbations, 
denoted as $\mathcal{A}_{\zeta}$, should be on the order of $\mathcal{O}(0.01)$ \cite{Yi:2022ymw}. 
However, the measurements of cosmic microwave background (CMB) anisotropies at large scales constrained the amplitude as $\mathcal{A}_{\zeta}= 2.1\times 10^{-9}$ \cite{Akrami:2018odb}. 
Consequently, to account for the signal detected by PTAs, 
a significant enhancement of approximately seven orders of magnitude in  primordial curvature perturbations is required at small scales. 
This enhancement can be achieved through inflation models incorporating a transitional ultra-slow-roll phase \cite{Martin:2012pe, Motohashi:2014ppa,Yi:2017mxs, Garcia-Bellido:2017mdw,Germani:2017bcs, Motohashi:2017kbs,Ezquiaga:2017fvi, Gong:2017qlj, Ballesteros:2018wlw,Dalianis:2018frf, Bezrukov:2017dyv, Kannike:2017bxn, Gao:2019sbz,Lin:2020goi,Lin:2021vwc,Gao:2020tsa,Gao:2021vxb,Yi:2020kmq,Yi:2020cut,Yi:2021lxc,Yi:2022anu,Zhang:2020uek,Pi:2017gih,Kamenshchik:2018sig,Fu:2019ttf,Fu:2019vqc,Dalianis:2019vit,Gundhi:2020zvb,Cheong:2019vzl,Zhang:2021rqs,Kawai:2021edk,Cai:2021wzd,Chen:2021nio,Zheng:2021vda,Karam:2022nym,Ashoorioon:2019xqc}. 

The energy density spectrum of GW background is usually parameterized as the power-law form  $h^2\Omega_\text{GW} = A_\text{GW} (k/k_\mathrm{ref})^{n_\text{GW}}$,
the index $n_\text{GW}=2/3$ for GWs from SMBHBs, 
$n_\text{GW}=3$ for GWs from domain walls before the peak frequency  \cite{Hiramatsu:2013qaa}.
For SIGWs in the infrared region $k\ll k_p$, the index has a universal log-dependent behavior,  $n_{\rm GW}(k,k_p) = 3-2/\ln(k_p/k)$ \cite{Cai:2019cdl,Yuan:2019wwo,Xu:2019bdp,Pi:2020otn}.  
The GW signals in NANOGrav 15-year data set and EPTA Data Release 2 (DR2) exhibit a characteristic of increasing energy density with the frequency,  
which is consistent with the universal behavior of the SIGWs in infrared regions.
Furthermore, SIGWs from three particular models give a better fit to the data \cite{NANOGrav:2023hvm}.
In order to check whether this is true for more general models,
we consider SIGWs from the primordial scalar power spectrum with the broken power-law form of index $n_1$ before the peak $k_p$.
For this model, in the infrared region $k\ll k_p$,
$n_\text{GW}=2n_1$ if $n_1<3/2$,
$n_{\rm GW}(k,k_p) = 3-3/\ln(k_p/k)$ if $n_1=3/2$, 
and $n_{\rm GW}(k,k_p) = 3-2/\ln(k_p/k)$ if $n_1>3/2$. 
The constant index $n_\text{GW}=2n_1$ for $n_1<3/2$ can represent a wide class of models, like the SMBHB model with $n_\text{GW}=2/3$ and domain walls with $n_\text{GW}=3$.
The index $n_\text{GW}$ is independent of $n_1$ if $n_1\ge 3/2$, 
so the energy density spectrum of SIGWs in the infrared region does not depend much on a particular model and can be used to check the evidence of SIGWs in a somewhat model-independent way. 
By using the universal behavior of SIGWs in the infrared region, 
we aim to provide an explanation for the signal observed by PTAs with SIGWs.

The organization of this paper is as follows. In Section \ref{ap:sigw_inf}, we first briefly review the calculation of SIGWs and then we derive the behavior of SIGWs in infrared regions.
The explanation of the PTAs data by the SIGWs in the infrared region is presented in Section \ref{sec:result},
and conclusions are drawn in Section \ref{sec:con}.

\section{SIGWs and infrared behavior} \label{ap:sigw_inf}
The large scalar perturbations originating from the primordial curvature perturbation generated during inflation can serve as a source to induce gravitational waves, during the radiation domination epoch. 
The power spectrum of SIGWs is
\cite{Kohri:2018awv,Lu:2019sti,Baumann:2007zm,Inomata:2016rbd,Espinosa:2018eve,Ananda:2006af}
\begin{equation}\label{Eph2}
\begin{split}
\mathcal{P}_{h}\left(k,\eta\right)=&4\int^{\infty}_{0}\text{d}v\int^{1+v}_{\left|1-v\right|}\text{d}u \left(\frac{4v^2-\left(1+v^2-u^2\right)^2}{4vu}\right)^2\\& \quad \quad \quad
\times I^{2}_{\mathrm{RD}}\left(u,v,x\right)\mathcal{P}_\zeta\left(vk\right)\mathcal{P}_\zeta\left(uk\right),
\end{split}
\end{equation}
where $x=k\eta$. 
 At late time, $k\eta \gg 1$, i.e., $x\rightarrow \infty$, the time average of ${I^{2}_{\mathrm{RD}}\left(u,v,x\rightarrow\infty \right)}$ is \cite{Kohri:2018awv}
\begin{equation}\label{IRD}
\begin{split}
 \overline{I_{\mathrm{RD}}^2(v,u,x\rightarrow \infty)} &= \frac{1}{2x^2} \left( \frac{3(u^2+v^2-3)}{4 u^3 v^3 } \right)^2 \\& \times \left\{\pi^2 (u^2+v^2-3)^2 \Theta \left( v+u-\sqrt{3}\right) \right. \\& \left. -\left(4uv-(u^2+v^2-3) \log \left| \frac{3-(u+v)^2}{3-(u-v)^2} \right| \right)^2 \right\}.
\end{split}
\end{equation}
The definition of the fractional energy density of SIGWs per logarithmic interval of ${k}$ is
 \begin{equation}\label{EGW0}
\Omega_{\mathrm{GW}}\left(\eta,k\right)=\frac{1}{24}\left(\frac{k}{\mathcal{H\left(\eta\right)}}\right)^2\overline{\mathcal{P}_h\left(k,\eta\right)}.
\end{equation}
Because the energy density of  SIGWs decays similarly to radiation, we can estimate the energy density of SIGWs today in terms of the present energy density of radiation, denoted as ${\Omega_{r,0}}$. And the relationship is \cite{Espinosa:2018eve}
\begin{equation}\label{EGW}
\Omega_{\mathrm{GW}}\left(\eta_0,k\right)=\Omega_{\mathrm{GW}}\left(\eta,k\right)\frac{\Omega_{r,0}}{\Omega_{r}\left(\eta\right)},
\end{equation}
where ${\eta}$ can be chosen at a generic time towards the end of the radiation domination era.

The GW signals detected in NANOGrav 15-year data set and EPTA DR2 exhibit a characteristic of increasing energy density with frequency. 
This characteristic is in line with the model-independent behavior of the energy density spectrum of  SIGWs in infrared regions.  
To explore the   SIGWs, 
several parameterized power spectra of primordial curvature perturbations were proposed \cite{Ananda:2006af,Saito:2008jc,Namba:2015gja,Nakama:2015nea,Nakama:2016enz,Bartolo:2018evs,Byrnes:2018txb,Orlofsky:2016vbd,Xu:2019bdp,Inomata:2016rbd,Garcia-Bellido:2017aan,Kohri:2018awv,Bartolo:2018rku, Cai:2018dig,Lu:2019sti,Unal:2018yaa,Inomata:2018epa,Germani:2018jgr,DeLuca:2019llr,Pi:2020otn}.
Among these, a straightforward parameterization is the broken power-law function,
which is capable of capturing the peak in the power spectrum that arises in inflationary model \cite{Bullock:1996at,Saito:2008em,Kadota:2015dza,Inomata:2016rbd,Ballesteros:2017fsr,Cicoli:2018asa,Cheng:2018qof,Fu:2019ttf,Bhaumik:2019tvl,Lin:2020goi,Dalianis:2018frf,Tada:2019amh,Fu:2019vqc,Xu:2019bdp,Yi:2020kmq,Yi:2022ymw, Inomata:2016rbd,Inomata:2017okj,Lu:2019sti}.

In this paper, we parameterize the power spectrum of the primordial curvature perturbation with the broken power-law form,
\begin{equation}\label{bk_ps}
\mathcal{P}_{\zeta}\left(k\right)=\begin{cases}
\mathcal{A}_{\zeta}\left(\frac{k}{k_p}\right)^{n_{1}},&{k\le k_p},\\
\mathcal{A}_{\zeta}\left(\frac{k}{k_p}\right)^{n_{2}},&{k>k_p}.
\end{cases}
\end{equation}
where $k_p$ is the peak scale and $\mathcal{A}_{\zeta}$ is the amplitude at the peak, and the power indices $n_1>0$ and $n_2<0$.

For simplicity of notation, we define the following function
\begin{equation}\label{f}
\begin{split}
f(u,v)=&\frac{x^2}{6}\left(\frac{4v^2-\left(1+v^2-u^2\right)^2}{4vu}\right)^2\overline{I^{2}_{RD}\left(u,v,x \rightarrow \infty \right)},
\end{split}
\end{equation}
 and divide the integral Eq. \eqref{EGW0} into three parts,
\begin{align}\label{GW}
\Omega_{\mathrm{GW}}(k)=&\int^{c_1}_{0}dv \int^{1+v}_{|1-v|}du f(u,v)\mathcal{P}_\zeta\left(vk\right)\mathcal{P}_\zeta\left(uk\right)\nonumber\\&+\int^{c_2}_{c_1}dv \int^{1+v}_{|1-v|}du f(u,v)\mathcal{P}_\zeta\left(vk\right)\mathcal{P}_\zeta\left(uk\right)\nonumber\\&
+\int^{\infty}_{c_2}dv \int^{1+v}_{|1-v|}du f(u,v)\mathcal{P}_\zeta\left(vk\right)\mathcal{P}_\zeta\left(uk\right),
\end{align}
with $c_1\ll1$ and $c_2\gg1$. Utilizing the mean value theorem in calculus, we can approximately express the first and last terms on the right-hand side of Eq. \eqref{GW} as follows
\begin{equation}
\int^{c_1}_{0}dv \int^{1+v}_{|1-v|}du f(u,v)\mathcal{P}_\zeta\left(vk\right)\mathcal{P}_\zeta\left(uk\right) \approx 2\int^{c_1}_{0}dv vf(1,v)\mathcal{P}_\zeta\left(vk\right)\mathcal{P}_\zeta\left(k\right),
\end{equation}
and 
\begin{equation}
\int^{\infty}_{c_2}dv \int^{1+v}_{|1-v|}du f(u,v)\mathcal{P}_\zeta\left(vk\right)\mathcal{P}_\zeta\left(uk\right)\approx
2\int^{\infty}_{c_2}dv f(v,v)\mathcal{P}_\zeta\left(vk\right)\mathcal{P}_\zeta\left(vk\right),
\end{equation}
respectively. 
In the limitation $v\leq c_1\ll1$, we have
\begin{equation}\label{f2}
f(1,v)\simeq \frac{1}{3}v^2, 
\end{equation}
and $1\ll  c_2 \leq v$, $f(v,v)$ behaves as
\begin{equation}\label{f1}
\begin{split}
f(v,v)\simeq& \frac{3}{4}\left[4+\pi^2-4\ln(4/3)+\left(\ln(4/3)\right)^2\right.\\&\left.-4\left(2-\ln(4/3)\right)\ln(v)+4\left(\ln v\right)^2\right]\frac{1}{v^4}.
\end{split}
\end{equation}

Using the approximate expressions \eqref{f2} and \eqref{f1}, we are able to calculate the energy density $\Omega_{\mathrm{GW}}$ of  SIGWs in the infrared regions, where $k$ is much smaller than the peak scale $k_p$, $k_p/k>c_2\gg1$.   With the Eqs. \eqref{bk_ps}, \eqref{GW}, \eqref{f2}, and \eqref{f1}, we can obtain the energy density $\Omega_{\mathrm{GW}}$ for $n_1\neq 3/2$ in the infrared regions,
\begin{equation}\label{left1}
\begin{split}
\frac{\Omega_{\mathrm{GW}}}{\mathcal{A}^2_{\zeta}}\simeq & \left(\frac{k}{k_p}\right)^{2n_1}\left[\frac{2}{3}\frac{c_1^{4+n_1}}{4+n_1}+g(n_1)
-c_2^{2n_1}\Pi(n_1, c_2)\right] \\& +\Pi(n_1, k_p/k) -\Pi(n_2, k_p/k),
\end{split}
\end{equation}
where the function $\Pi$ is 
\begin{equation}
\Pi(n,x) =\frac{3}{2}\frac{1}{(2n-3)^3 x^{3}}
\big[ A(n)+B(n)\ln x +4(3-2n)^2 \ln^2 x \big],
\end{equation}
and 
\begin{align}
A(x) &=20+9\pi^2-24\ln\frac{4}{3}+9 \ln^2\frac{4}{3}\nonumber \\ &-4x \left(8+3\pi^2-10\ln\frac{4}{3}+3 \ln^2\frac{4}{3}\right)\nonumber \\
&+4x^2\left(4+\pi^2-4\ln\frac{4}{3}+ \ln^2\frac{4}{3}\right),\\
B(x) &=4(-3+2x)\left[4+x\left(-4+2\ln\frac{4}{3}\right)-3\ln\frac{4}{3}\right],
\end{align}
and
\begin{equation}
g(x)=\int^{c_2}_{c_1}dv\int^{1+v}_{|1-v|}duf(u,v)u^{x}v^{x}.
\end{equation}
Note that $g$ is a finite constant once we fix its argument.

Similarly, for $n_1=3/2$, we have
\begin{equation}\label{left2}
\begin{split}
\frac{\Omega_{\mathrm{GW}}}{\mathcal{A}^2_{\zeta}}\simeq &  \left(\frac{k}{k_p}\right)^{2n_1}\left[\frac{2}{3}\frac{c_1^{4+n_1}}{4+n_1}+g(n_1)\right]\\&
+\frac{3}{2}\left(\frac{k}{k_p}\right)^3\left[\Delta(k_p/k) -\Delta(c_2)\right] -\Pi(n_2, k_p/k).
\end{split}
\end{equation}
where the function $\Delta$ is 
\begin{equation}
    \Delta(x)= \left[\pi^2+\left(-2+\ln\frac{4}{3}\right)^2\right]\ln x 
    +2\left(-2+\ln\frac{4}{3}\right) \ln^2 x +\frac{4}{3} \ln^3 x.
\end{equation}
By utilizing the approximate expressions of SIGWs given in Eq. \eqref{left1} and \eqref{left2}, we can investigate the behavior of SIGWs in the infrared region.

For the case where $n_1<3/2$, the first term of \eqref{left1} is dominated  in infrared regions, we have \cite{Xu:2019bdp}
\begin{equation}
\Omega_{\mathrm{GW}}(k) \propto k^{2n_1},
\end{equation}
which means that the power index of SIGWs is always two times the index of the power spectrum for $n_1<3/2$.

For $n_1>3/2$, the second and last terms on the right-hand side of Eq. \eqref{left1} are dominant in the infrared regions. Both of these terms exhibit similar scaling behavior with respect to the wavenumber. We can obtain
\begin{equation}
n_{\mathrm{GW}}=\frac{d\ln\Omega_{\mathrm{GW}}(k)}{d\ln k}= 3-2/\ln(k_p/k),
\end{equation}
Note that this log-dependent behavior can be derived from more general primordial curvature power spectrum and is universal for narrow spectrum \cite{Yuan:2019wwo}.

For $n_1=3/2$, the second term in the right-hand side of Eq. \eqref{left2} is dominated,  we have
\begin{equation}
n_{\mathrm{GW}}=\frac{d\ln\Omega_{\mathrm{GW}}(k)}{d\ln k}=3-3/\ln(k_p/k).
\end{equation}

In this paper,  we  use a power-law function to parametrize $\Omega_{\text{GW}}$ in the infrared regions,
\begin{equation}\label{para:pl}
h^2 \Omega_{\mathrm{GW}} = A_\mathrm{GW} \left(\frac{k}{k_{0}}\right)^{n_{\mathrm{GW}}},
\end{equation}
where $k_0$ is an arbitrary scale. Although, the exact expression of the amplitude of $\Omega_{\text{GW}}$ may depend on the wavenumber $k$; however,  in the infrared regions, the amplitude can be approximated as a constant. Hence, the power-law function parameterization \eqref{para:pl} can be applied within the infrared regions. For GWs with nanohertz  frequencies, we choose $k_0 = k_{\rm ref}$, where $k_{\rm ref}$ is the scale  corresponding to the  frequency of $f_{\rm ref}=1~{\rm yr} ^{-1}$. 
The index $n_{\mathrm{GW}}$ is
\begin{equation}\label{leftomega2}
n_{\mathrm{GW}} = \begin{cases}
2n_1, & n_1 < \frac{3}{2},\\
3-3/ \ln(k_p/k), & n_1=\frac{3}{2},\\
3-2/ \ln(k_p/k), & n_1>\frac{3}{2}.
\end{cases}
\end{equation}
From the above results, if $n_1<3/2$, then the index $n_\text{GW}=2n_1$ is a constant;
but when $n_1\ge 3/2$, the index $n_\text{GW}$ is independent of $n_1$
and it has the universal log-dependent behavior. 
Although the above results derive from the power-law parametrized primordial power spectrum, 
the index of the energy density parameter of SIGWs scaling as 3 along with a log-dependent term is universal and independent of model, which has been proven in Refs. \cite{Yuan:2019wwo,Cai:2019cdl}.
As discussed above, the constant index $n_\text{GW}=2n_1$ for $n_1<3/2$ can be derived from several models, e.g., $n_\text{GW}=2/3$ for the SMBHB model. 
The log-dependence index for $n_1>3/2$ is valid for a wide class of models. 
Therefore, the parametrization \eqref{para:pl} with the index \eqref{leftomega2} is somewhat model independent in this sense
and the discussion based on the parametrization \eqref{para:pl} and \eqref{leftomega2} is also independent of SIGW model.

\section{The results}\label{sec:result}

In this section, we perform a Bayesian analysis on the combination of the NANOGrav 15-year data set and EPTA DR2 to investigate the parameters of SIGWs described by Eq. \eqref{para:pl}.  
We use the 14 frequency bins of NANOGrav 15-year data set \cite{NANOGrav:2023gor, NANOGrav:2023hvm} and 9 frequency bins of EPTA DR2 \cite{Antoniadis:2023ott, Antoniadis:2023zhi},  
to fit the posterior distributions of the parameters. The analysis was performed using the  \texttt{Bilby} code \cite{Ashton:2018jfp}, 
employing the  \texttt{dynesty} algorithm for nested sampling \cite{NestedSampling} with 1000 live points (${\rm nlive} = 1000$).  
The log-likelihood function was derived by evaluating the energy density of the SIGWs at the 23 specific frequency bins. 
We then calculate the sum of the logarithm of the probability density functions from 23 independent kernel density estimates \cite{Moore:2021ibq}. 
Consequently, the likelihood function can be expressed as 
\begin{equation}
\ln \mathcal{L}(\mathbf{\Theta})=\sum_{i=1}^{23} \ln \mathcal{L}_i\left(\Omega_{\mathrm{GW}}\left(f_i, \mathbf{\Theta} \right)\right),
\end{equation}
where $\boldsymbol{\Theta}$ is the collection of parameters presented in the parameterization  \eqref{para:pl}. These parameters and their priors are shown in Table~\ref{tab:prior_posterior}.

\begin{table*}[t]
\centering
\footnotesize
\tabcolsep 18pt
\begin{tabular*}{0.9\textwidth}{l|llll}
\hline
\hline
Model & Parameters & Prior & Posterior & $\ln \mathcal{B}$ \\
\hline 
\multirow{2}{*}{$n_1<3/2$} \quad 
 &$\text{log}_{10}A_\mathrm{GW}$ & $U[-9, -5]$ & $ -7.12^{+0.18}_{-0.21}$ &\multirow{2}{*}{$9.1$} \\
 &$n_1$       & $U[0,1.5]$    & $ 1.03^{+0.14}_{-0.15}$   & \\
\hline
\multirow{2}{*}{$n_1=3/2$} 
& $ \text{log}_{10}A_\mathrm{GW}$  & $U[-9, -5]$ & $-6.98^{+0.16}_{-0.22}$&\multirow{2}{*}{ $8.4$} \\
& $\text{log}_{10}(k_p/\text{Mpc}^{-1})$ & $U[6,11]$ & $8.50^{+1.10}_{-0.66}$ &\\
\hline 
\multirow{2}{*}{$n_1>3/2$}
&$\text{log}_{10}A_\mathrm{GW}$ & $U[-9, -5]$ & $-6.87^{+0.14}_{-0.17}$&\multirow{2}{*}{ $7.5$}\\
&$\text{log}_{10}(k_p/\text{Mpc}^{-1})$ & $U[6,11]$ & $8.40^{+1.30}_{-0.64}$ & \\
\hline 
\text{SMBHB} & $\rm log_{10}A_\mathrm{GW}$ & $U[-9, -5]$ & $-8.19^{+0.14}_{-0.08}$ &~ $0$\\
\hline 
\hline 
\end{tabular*}
\caption{The priors, maximum posterior values, 1-$\sigma$ credible interval bounds of posteriors and Bayes factor for the parametrization \eqref{para:pl}  using NANOGrav 15-year data set and EPTA DR2.
We set the  SMBHB  model as the fiducial model.}
\label{tab:prior_posterior}
\end{table*}

\begin{figure}[H]
  \centering
  \includegraphics[width=0.8\textwidth]{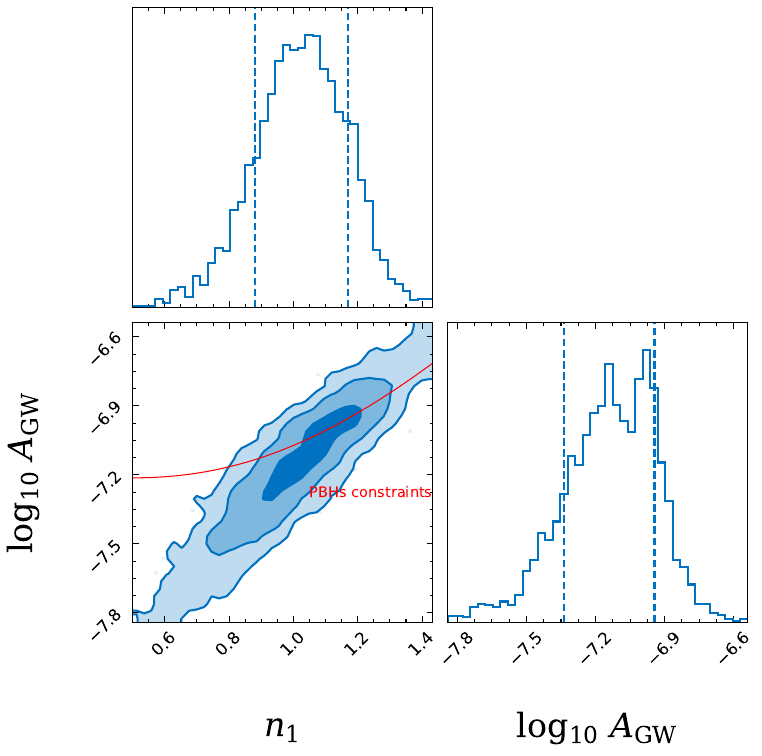}
  \caption{The posteriors on the parameters in  Eq. \eqref{para:pl} with $n_1<3/2$ from the NANOGrav 15-year data setand EPTA DR2. The red line represents the constraints from the PBH observational data from the EROS-2 project \cite{EROS-2:2006ryy}, and the allowed regions lie below the red line.}\label{pic:le}
  \end{figure}
  
\begin{figure}[H]
  \centering
  \includegraphics[width=0.8\textwidth]{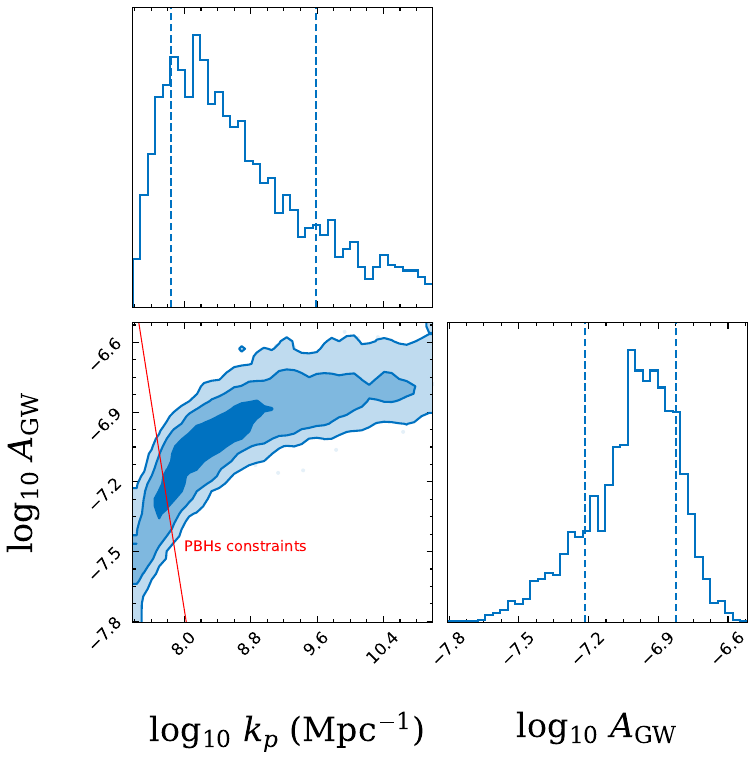}
  \caption{The posteriors on the parameters in  Eq. \eqref{para:pl} with $n_1=3/2$ from the NANOGrav 15-year data set  and EPTA DR2. The red line represents the constraints from the PBH observational data from the EROS-2 project \cite{EROS-2:2006ryy}, and the allowed regions lie below the red line.}
  \label{pic:eq}
  \end{figure}
  
  \begin{figure}[H]
  \centering
  \includegraphics[width=0.8\textwidth]{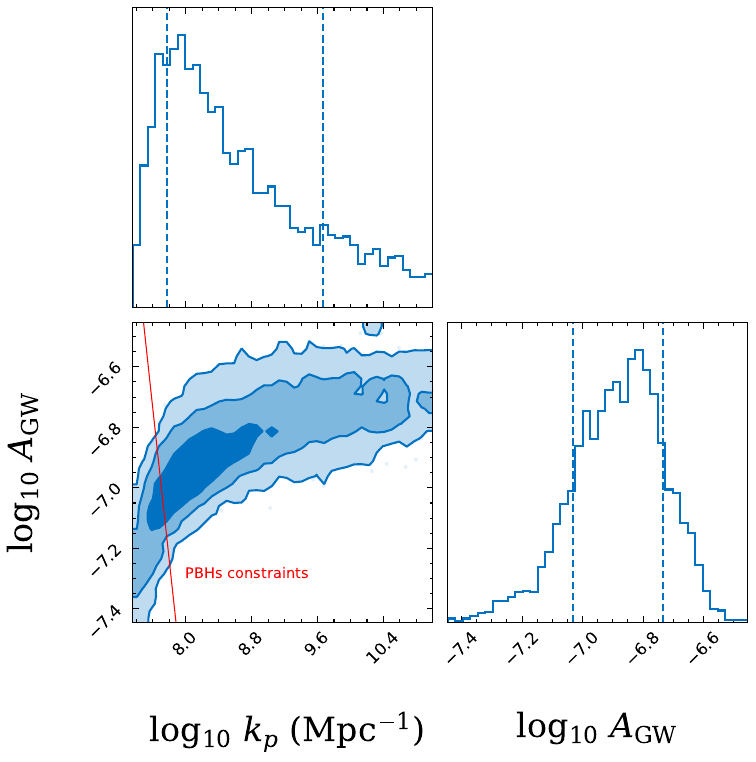}
  \caption{The posteriors on the parameters in Eq. \eqref{para:pl} with $n_1>3/2$ from  the NANOGrav 15-year data set  and EPTA DR2. The red line represents the constraints from the PBH observational data from the EROS-2 project \cite{EROS-2:2006ryy}, and the allowed regions lie below the red line.}\label{pic:lg}
  \end{figure}

The resulting posterior distributions for the model \eqref{bk_ps} with $n_1 < 3/2$, $n_1 = 3/2$, and $n_1 > 3/2$ are shown in Figs. \ref{pic:le}, \ref{pic:eq}, and \ref{pic:lg}, respectively.  
The mean values and 1-$\sigma$ confidence intervals for the parameters are summarized in Table \ref{tab:prior_posterior}. 
The frequency bins $f_{\rm bin}$ of the NANOGrav 15-year data set and EPTA DR2 satisfy the condition $f_{\rm bin} \lesssim f_p/10$ Hz,  
where $f_p =5.8\times 10^{-7}$ Hz is the mean value of the peak scale $\log_{10}(k_p/\mathrm{{Mpc}^{-1}}) =8.5$, 
ensuring that the scales we focus on satisfy $k \ll k_p$. 
The results guarantee that the frequencies in the data set fall within the infrared region and fulfill the necessary infrared condition. 
Comparing the interpretation of the detected signal as a stochastic background from SMBHBs with $n_\text{GW}=2/3$, the Bayesian analysis yields support in favor of 
SIGWs with respective Bayes factors of $\ln \mathcal{B}= 9.1$, $\ln \mathcal{B}= 8.4$, and $\ln \mathcal{B} = 7.5$ for the case with $n<3/2$, $n=3/2$, and $n>3/2$, respectively.

\begin{figure}
\centering
\includegraphics[width=0.8\textwidth]{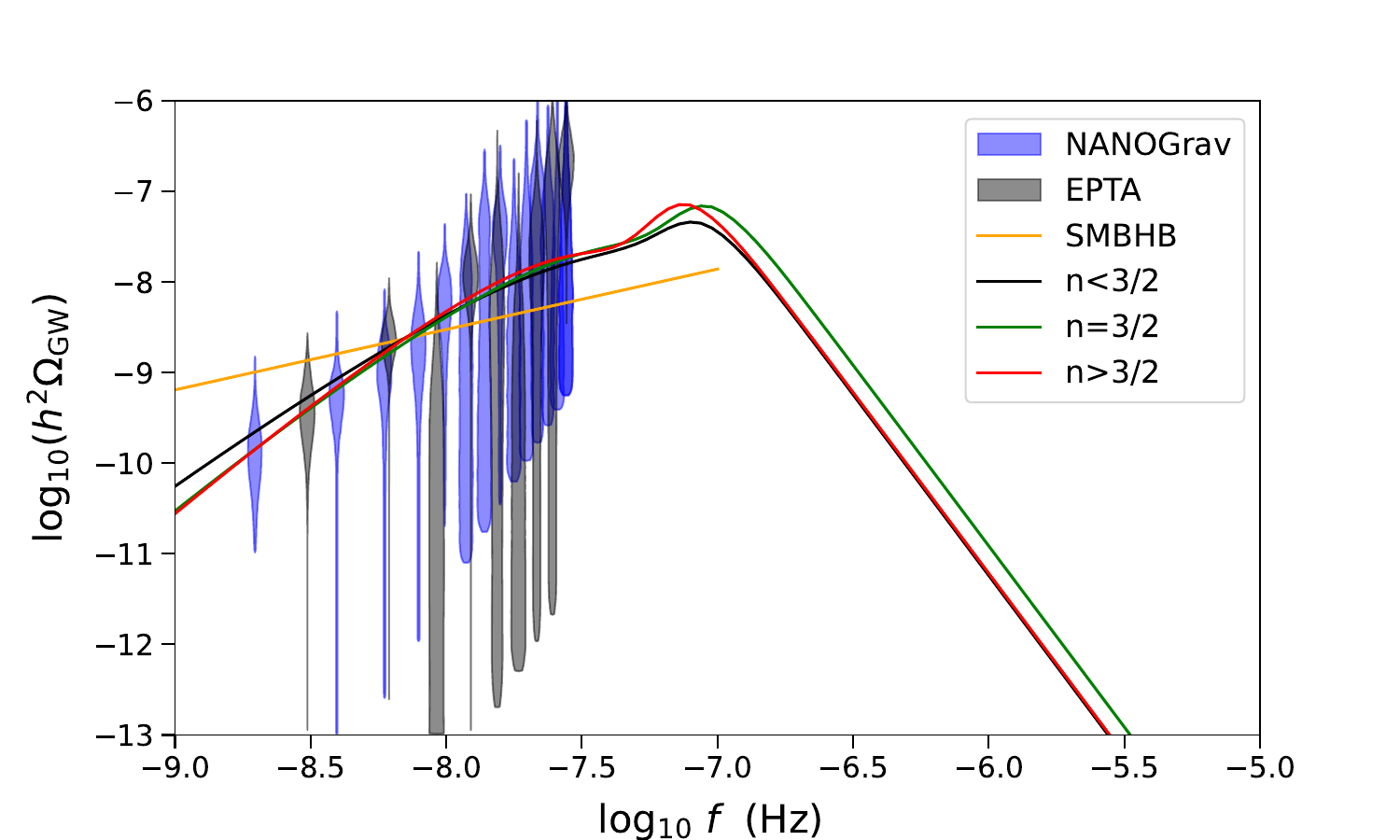}
\caption{The energy density of SIGWs  based on model \eqref{bk_ps}, utilizing the parameter values obtained from the best-fit parameter values of model \eqref{para:pl} as shown in Table \ref{tab:prior_posterior}. The black, green, and red curves represent the energy density of SIGWs with $n_1<3/2$, $n_1=3/2$, and $n_1>3/2$, respectively.  The orange line denotes the energy density of the gravitational wave from the scenario of SMBHB. 
The blue and grey violins represent the EPTA DR2 and NANOGrav 15-year data set, respectively.}\label{pic:compara}
\end{figure}

From the best-fit parameter values of the SIGWs parameterization \eqref{para:pl} displayed in Table~\ref{tab:prior_posterior}, we can determine the corresponding best-fit parameter values of the primordial curvature power spectrum \eqref{bk_ps}. By using these best-fit parameter values, we calculate the energy density of the SIGWs for all frequencies, as shown in Fig. \ref{pic:compara}.
The results show that the infrared parts of SIGWs from primordial curvature perturbations with different pow-law index all fit the data better than GW background from SMBHBs.
In the case with the constant index $n_\text{GW}$, 
the data prefers a bluer spectrum with $n_\text{GW}\approx 2$ which is consistent with those found in \cite{Moore:2021ibq,NANOGrav:2023gor,Zic:2023gta,Reardon:2023gzh,Antoniadis:2023ott,NANOGrav:2023hvm}.
For a more bluer spectrum with $n_\text{GW}\approx 3$,
the log-dependent index \eqref{leftomega2} is needed.
Even though the characteristic log-dependence existed in the index $n_\text{GW}$ for SIGWs,
the current data is unable to distinguish the GW spectrum with the characteristic log-dependence from those with a constant as shown in Fig. \ref{pic:compara} and also seen from the Bayes factor displayed in Table \ref{tab:prior_posterior}.

Accompanying the generation of SIGWs, the large scalar perturbations can produce primordial black holes through gravitational collapse of overdensity regions at the horizon reentry during radiation domination.  
The current fractional energy density of PBHs to dark matter is 
\begin{equation}
\label{pbh:fpbheq1}
\begin{split}
Y_{\text{PBH}}(M)
= \frac{\beta(M)}{3.94\times10^{-9}}\left(\frac{\gamma}{0.2}\right)^{1/2}
\left(\frac{g_*}{10.75}\right)^{-1/4} \times \left(\frac{0.12}{\Omega_{\text{DM}}h^2}\right)
\left(\frac{M}{M_\odot}\right)^{-1/2},
\end{split}
\end{equation}
where $M_{\odot}$ is the solar mass, $\gamma= 0.2$ \cite{Carr:1975qj}, $g_*$ is the effective degree of freedom at the formation time, 
and the current energy density parameter of dark matter $\Omega_{\text{DM}}h^2=0.12$ \cite{Planck:2018vyg}. 
The PBH mass fraction at the formation time can be obtained by the Press-Schechter theory, 
\begin{equation}\label{beta:pro}
  \beta=\int_{\delta_c}^{\infty}P(\delta) d\delta,
\end{equation}
where $P(\delta)$ is the probability distribution function (PDF) of density contrast $\delta$, and $\delta_c$ is the threshold for the formation of PBHs. Here we choose $\delta_c=0.45$.
The relation between the PBH mass $M$ and the scale $k$ is \cite{Gong:2017qlj}
\begin{equation}
\label{mkeq1}
M(k)=3.68\left(\frac{\gamma}{0.2}\right)\left(\frac{g_*}{10.75}\right)^{-1/6}
\left(\frac{k}{10^6\ \text{Mpc}^{-1}}\right)^{-2} M_{\odot}.
\end{equation}    

For primordial curvature power spectrum parameterized by the broken power-law model \eqref{bk_ps}, 
we can obtain upper limits on the amplitude $\mathcal{A}_\zeta$ based on observational constraints on PBHs. 
To derive the constraints, we adopt the parameter values for the broken power-law model as follows:  $\log_{10}(k_p/\mathrm{Mpc}^{-1}) =7.7$ and $n_2=-2$ for the scenario with $n_1<3/2$, $n_2=-2$ for the scenario with $n_1=3/2$, and $n_1=2$ and $n_2=-2$ for the scenario with $n_1>3/2$. The abundance of  PBHs generated by the power spectrum of curvature perturbations, peaking at a scale approximately $ \mathcal{O}(10^8) ~\rm{Mpc}^{-1}$, is constrained by the EROS-2 project \cite{EROS-2:2006ryy}.  The upper limits of the amplitude $\mathcal{A}_{\zeta}$ can be obtained by ensuring that the PBH abundance remains within the observational constraints set by the EROS-2 project. Furthermore, the constraints on the amplitude of SIGWs can be obtained from the constraints on the primordial curvature power spectrum using the formulas displayed in section \ref{ap:sigw_inf},  and these constraints are represented by the red lines in Figs. \ref{pic:le}, \ref{pic:eq}, and \ref{pic:lg}.   The parameter regions below the red lines are allowed by the PBH constraints, providing valuable information about the viability of the model under consideration. 

\section{conclusion}\label{sec:con}
The GW signal detected by NANOGrav, PPTA, EPTA, and CPTA collaborations can be explained by SIGWs. 
The energy density spectrum of SIGWs from primordial scalar perturbations with broken power-law form can be parameterized as power-law form in the asymptotic regions.
In the infrared regions $k\ll k_p$, 
$h^2 \Omega_{\mathrm{GW} } = A_\mathrm{GW}\left(k/k_{\rm ref}\right)^{n_{\mathrm{GW}}}$, where the index $n_{\mathrm{GW}}$ takes different values depending on the range of $n_1$: $n_{\mathrm{GW}} = 2 n_1$ for $n_1<3/2$, $n_{\mathrm{GW}} = 3-3/ \ln(k_p/k)$ for $n_1=3/2$, and $n_{\mathrm{GW}} =3-2/ \ln(k_p/k)$ for $n_1>3/2$.
Even though the result is derived from a particular model,
it is actually applicable for more general models. 
The constant index $n_\text{GW}=2n_1$ for $n_1<3/2$ can represent a wide class of models, like the SMBHB model with $n_\text{GW}=2/3$ and domain walls with $n_\text{GW}=3$.
The log-dependence index for $n_1>3/2$ is valid for a wide class of models. 
Therefore, the parametrization \eqref{para:pl} with the index \eqref{leftomega2} is somewhat model independent
and the result based on the parametrization \eqref{para:pl} and \eqref{leftomega2} is also independent of SIGW model.

For the PTA signals, we demonstrate that it can be effectively explained by the near-model-independent behavior of  SIGWs in the infrared regions. 
Through Bayesian analysis, we have identified the parameter space that can adequately account for the combination of NANOGrav 15-year data set and EPTA DR2.  
For the case with constant power index $n_\text{GW}=2n_1$, 
the mean values and one-sigma confidence intervals are  $\log_{10} A_\mathrm{GW} = -7.12^{+0.18}_{-0.21}$ and $n_1 = 1.03^{+0.14}_{-0.15}$. 
For the case with the log-dependence index $n_\text{GW} =3-3/ \ln(k_p/k)$, 
the mean values and one-sigma confidence intervals are   $\log_{10} A_\mathrm{GW} = -6.98^{+0.16}_{-0.22}$ and $\log_{10} (k_p/ {\rm Mpc}^{-1}) = 8.50^{+1.10}_{-0.66}$. 
For the case with the log-dependence index $n_\text{GW} =3-2/ \ln(k_p/k)$, 
the mean values and one-sigma confidence intervals are  $\log_{10} A_\mathrm{GW} = -6.87^{+0.14}_{-0.17}$ and $\log_{10} (k_p/ {\rm Mpc}^{-1}) = 8.40^{+1.30}_{-0.64}$. 
The last two scenarios provide an alternative explanation where the index of the SIGWs is scale-dependent.  
Comparing with the interpretation of the detected signal as a stochastic background from SMBHBs, 
the Bayes factors for the above three cases are $\ln \mathcal{B}= 9.1$, $\ln \mathcal{B}= 8.4$, and $\ln \mathcal{B} = 7.5$, respectively. 
Therefore, our results provide tentative support for SIGWs. 
Furthermore, we provide constraints on the parameters of the SIGW models with observational data from PBHs, 
revealing that about half of the parameter regions of the SIGWs allowed by PTAs are ruled out by the PBH constraints for the particular model. The overproduction of PBHs is a significant issue when considering SIGWs as an explanation for the PTA signal. To address this issue, incorporating non-Gaussianities may offer a promising approach, as discussed in Refs. \cite{Franciolini:2023pbf, Liu:2023ymk}.

In conclusion, our findings contribute to a comprehensive understanding of the PTA signals and emphasize the potential significance of SIGWs in explaining the observed data. 
However, it is important to note that the current data is unable to distinguish whether the power index of the energy density of SIGWs in infrared regions remains constant or exhibits variation. 
Further investigations and more precise measurements will be required to discern the true nature of SIGWs and their implications for our understanding of the Universe.

\begin{acknowledgments}
This work was supported by the National Key Research and Development Program of China (Grant No. 2020YFC2201504),
and the National Natural Science Foundation of China (Grant No. 12175184).
Z. Yi was supported by the National Natural Science Foundation of China (Grant No. 12205015) and the supporting fund for young researcher of Beijing Normal University (Grant No. 28719/310432102). F. Zhang was supported by the National Natural Science Foundation of China (Grant No. 12305075). Z. Yi thanks Lang Liu for  useful discussions.
\end{acknowledgments}


%

\end{document}